\documentstyle[prd,aps,floats,graphicx]{revtex}
\begin{document}
\draft

\twocolumn[\hsize\textwidth\columnwidth\hsize\csname
@twocolumnfalse\endcsname

\title{\bf Applications of scalar attractor solutions to Cosmology}

\author{S.C.C. Ng \cite{sccng}, ~N.J. Nunes \cite{njnunes} and ~F. Rosati 
\cite{frosati}}

\address{ \cite{sccng}Department of Physics and Mathematical Physics, University of
Adelaide, Adelaide, S.A. 5005, Australia}
\address{ \cite{njnunes}Centre for Theoretical Physics, University of Sussex,
Falmer, Brighton BN1 9QJ, United Kingdom }
\address{ \cite{frosati}Dipartimento di Fisica `Galileo Galilei', 
Universit\`{a} di Padova, via Marzolo 8, 35131 Padova, Italy \\ 
and INFN - Sezione di Padova, via Marzolo 8, 35131 Padova, 
Italy }

\date{\today}

\maketitle

\begin{abstract} 
We develop a framework to study the phase space of a system consisting 
of a scalar 
field rolling down an arbitrary potential with 
varying slope and a background fluid, in a cosmological setting.
We give analytical approximate solutions of the field evolution and
discuss applications of its features to the issues of quintessence,
moduli stabilisation and quintessential inflation. 
\end{abstract}
\pacs{PACS numbers: $98.80.$Cq \hspace*{1cm} astro-ph/xxx 
\hspace*{1cm}  SUSX-TH/01-030 \hspace*{1cm}  DFPD-01/TH/20 }
\vskip2pc]

%
%%%%%%%%%%%%%%%%%%%%%%%%%%%%%%%%%%%%%%%%%%%%%%%%%
\section{Introduction}
%%%%%%%%%%%%%%%%%%%%%%%%%%%%%%%%%%%%%%%%%%%%%%%%%%%
Scalar fields play a central role both in particle physics and 
cosmology. For example, in the Standard Model of particle physics,
the Higgs boson generates particle masses through
a mechanism of symmetry breaking. Supersymmetric
models, believed to solve the hierarchy problem and to 
suggest a grand unification scale, add a whole new set of scalar
(and fermionic) partners to the particles of the Standard Model.
In string theories, the only physical constant is the string tension
being all the other constants generated
dynamically through scalar fields, the moduli. 

In cosmology, a scalar field,
the inflaton, has been suggested to be responsible for an
early inflationary period in the history of the universe. This scenario
can account for the extreme flatness and homogeneity of our
Universe \cite{guth}. More recently, measurements of the apparent
magnitude--redshift relation of type Ia supernovae (SnIa) support indications
given by the combination of cosmic microwave background radiation (CMB),
galaxy clusters and light elements abundances measurements, that the
Universe is presently undergoing a period of accelerated expansion \cite{data}.
Once again, the existence of a scalar field rolling down a potential has
been suggested as an explanation for this late inflationary period. 
The scalar 
field is usually called ``quintessence'' \cite{zlatev}. However, whether the
inflaton and quintessence could be the same field, remains an 
interesting open question.
  
Considering these and other examples of applications of dynamical scalar fields
in physics, it seems important to take a closer look to the features of their
potentials. 

For instance, in the issue of quintessence, if the potentials have attractor 
solutions ({\it i.e.} the late time dynamics is independent of the
initial conditions) then there is a chance to weaken the fine tuning
problem associated with a cosmological constant term in Einstein's equations
\cite{stein}.
In the same way, when allowing for dynamical evolution of the moduli
fields, the existence of attractor
solutions opens up a larger region of initial conditions for which the 
fields can have successful stabilisation at their vacuum expectation 
value \cite{barreiro98a,huey}.

The aim of this work is to extend the results of 
\cite{scalcosmo,scherrer,copeland,cintia}
to more generic type of potentials, motivated by the issues of quintessence, 
moduli stabilisation and quintessential inflation and to give
analytical support to some of
the conclusions in \cite{macorra}.

%%%%%%%%%%%%%%%%%%%%%%%%%%%%%%%%%%%%%%%%%%%%%%%%
\section{Setup} 
%%%%%%%%%%%%%%%%%%%%%%%%%%%%%%%%%%%%%%%%%%%%%%%%
We consider a spatially--flat Friedmann--Robertson--Walker Universe 
containing a scalar field $\phi$ with potential $V(\phi)$, and a 
barotropic fluid with equation of state $p_B = (\gamma-1)\rho_B$, 
where $\gamma$ is a constant ({\it e.g.} $\gamma = 4/3$ for 
radiation and $\gamma = 1$ for matter). The governing equations of
motion are, 
\begin{eqnarray}
\dot{H} &=&- \frac{\kappa^2}{2} \left( \gamma \rho_B + \dot{\phi}^2 
\right) \,, \\
\dot{\rho}_B &=& -3 \gamma H \rho_B \,, \\
\ddot{\phi} &=& - 3 H \dot{\phi} - \frac{d V}{d \phi}  \, ,
\end{eqnarray}
subject to the Friedmann constraint
\begin{equation}
H^2 = \frac{\kappa^2}{3} \left(\rho_B + \frac{1}{2}\dot{\phi}^2 + V
\right) \,,
\end{equation}
where $\kappa^2 = 8 \pi G$ and dots denote derivatives with respect to time. 
The energy density and pressure of a
homogeneous scalar field are given by $\rho_{\phi} = 
\dot{\phi}^2/2 + V$ and $p_{\phi} = \dot{\phi}^2/2 - V$, respectively.

Following \cite{copeland}, we define the variables,
\begin{eqnarray}
x \equiv \frac{\kappa \phi'}{\sqrt{6}} , \hspace{1cm} 
y^2 \equiv \frac{\kappa^2 V}{3 H^2},
\end{eqnarray}
where a prime denotes a derivative with respect to the logarithm
of the scale factor $a$, $N \equiv \ln a$.

The effective equation of state for the scalar field at any point yields,
\begin{equation}
\gamma_{\phi} \equiv \frac{\rho_{\phi} + p_{\phi}}{\rho_{\phi}} = 
\frac{\dot{\phi}^2}{\dot{\phi}^2/2 + V} = \frac{2x^2}{x^2+y^2} ,
\end{equation}
constrained between $0 \le \gamma_{\phi} \le 2$.
In terms of these new variables the equations of motion read:
\begin{eqnarray}
\label{dynamics}
x' &=& -3x + \lambda \sqrt{\frac{3}{2}}y^2 + \frac{3}{2}x~[2x^2 +
       \gamma(1-x^2-y^2)] \nonumber \,, \\
y' &=& -\lambda\sqrt{\frac{3}{2}}x y + \frac{3}{2}y~[2x^2 +
       \gamma(1-x^2-y^2)] \,, \\
\lambda' &=& -\sqrt{6} \lambda^2 (\Gamma -1)x \nonumber \,,
\end{eqnarray}
where we have defined 
\begin{eqnarray}
\lambda \equiv - \frac{1}{\kappa V} \frac{d V}{d \phi} \,, \hspace{1cm}
\Gamma \equiv V \frac{d^2 V}{d \phi^2} / \left( \frac{d V}{d \phi} \right)^2 .
\end{eqnarray}
See \cite{macorra} and \cite{stein} respectively, where these definitions
were first introduced. Note that both $\lambda$ and $\Gamma$ are, in
general, $\phi$ (and thus, time) dependent.

The contribution of the scalar field to the total energy density,
 $\Omega_{\phi} \equiv \kappa^2 \rho_{\phi}/3 H^2 = 
 x^2 + y^2$ is bounded, $ 0 \leq \Omega_{\phi} \leq 1$, if
$\rho \geq 0$. Hence, the evolution of the system is completely
described by trajectories within the unit circle. Moreover, since
the system is symmetric under the reflection $(x,y) \rightarrow
(x,-y)$ and time reversal $ t \rightarrow -t$, we only consider 
the upper half disc, $y \geq 0$ in what follows.

\begin{figure}[ht!]
\includegraphics[height=6cm,width=8cm]{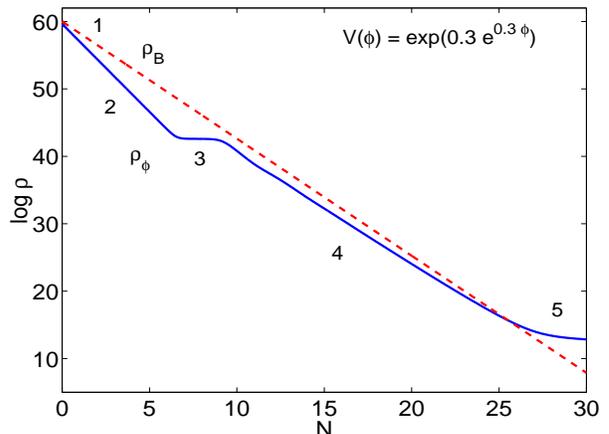}
\caption{ \label{fig1} Evolution of the scalar field energy density
$\rho_{\phi}$ in a radiation background fluid $\rho_B$ for a double exponential
potential. Regions 2 to 5 represent respectively, kination, frozen
field, evolution in the attractor and scalar field domination.}
\end{figure}

\begin{figure}[ht!]
\includegraphics[height=6cm,width=8cm]{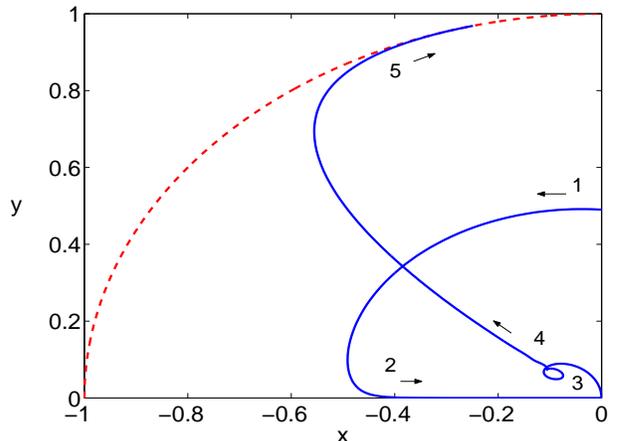}
\caption{ \label{fig2} The scalar field dynamics in the phase space
diagram for a double exponential potential. Regions 1 to 5 are the
same as in Fig. \ref{fig1}. The dashed line represents the unit
circle $x^2+y^2 = 1$.}
\end{figure}

For a generic scalar potential, one can identify up to five regions 
in the phase space
diagram $(x,y)$. As an example,
in Fig. \ref{fig1} and Fig. \ref{fig2} we show an exact numerical
solution of the evolution of
the system for a double exponential potential. In these figures,
region 1 represents a regime in which the
potential energy rapidly converts into kinetic energy; in region 2,
the kinetic energy is the dominant contribution to the total energy
density of the scalar field (``kination''); in region 3, the field
remains nearly constant until the attractor
solution is reached (``frozen field''); in region 4 the field evolves 
into the attractor
solution, where the ratio of the kinetic to potential energy is a
constant or slowly varying; and in region 5 the potential energy
becomes important, the scalar field dominates and drives the
dynamics of the Universe.

In section III we will briefly discuss regions 1, 2, and 3. 
However, in this paper we will be mostly concerned 
with regions 4 and 5 of the evolution since (as it will be shown)
they correspond 
to stable solutions yet not having true critical
points associated with them. 
As we will see, this feature plays an important role in many cosmological 
phenomena.
    

%%%%%%%%%%%%%%%%%%%%%%%%%%%%%%%%%%%%%%%%%%%%%%%%%%%%%%
\section{Early Evolution}
%%%%%%%%%%%%%%%%%%%%%%%%%%%%%%%%%%%%%%%%%%%%%%%%%%%%%%

For a wide range of potentials (including the potential in Figs. \ref{fig1}
and \ref{fig2}), when $\phi$ is high up the potential slope at the beginning
of the cosmological evolution, then $\lambda$ is large initially. 
The behaviour of the trajectory close to 
the $\lambda=\infty$ surface can be deduced by continuity 
from the $\lambda=\infty$ solution, even though the latter is not physical.

We bring the plane $\lambda=\infty$ to a finite distance from the $\lambda=0$
plane by the transformation
\begin{equation}
\lambda={\delta\over1-\delta} \ , \hspace{1cm}  0\le\delta\le1 \ ,
\end{equation}
so that the $\delta=0$ surface corresponds to the $\lambda=0$ surface and
the $\delta=1$ surface corresponds to $\lambda=\infty$. On the plane 
$\delta=1$, we find
\begin{eqnarray}
\label{dxdtau} \frac{dx}{d\tau} = \sqrt{3\over2}y^2 \,, \hspace{1cm} 
\label{dydtau} \frac{dy}{d\tau} = -\sqrt{3\over2}xy \, 
\end{eqnarray}
where $\tau$ is a new time coordinate defined by $d\tau=\lambda dN$ \cite{js}. 

Solving Eqs. (\ref{dxdtau}), we obtain,
\begin{eqnarray}
\label{xtanh} x&=&A\tanh \left[\sqrt{3\over2}(\tau-\tau_0)\right] \ , \\
\label{ysech} y&=&A\mbox{ sech}\left[\sqrt{3\over2}(\tau-\tau_0)\right] \ ,
\end{eqnarray}
where $A$ and $\tau_0$ are arbitrary constants. 

As shown in 
Fig. \ref{fig2}, the trajectory for region 1 is nearly circular, well 
described by Eqs. (\ref{xtanh}) and (\ref{ysech}). For $\lambda$ very 
large, the scalar field potential energy turns into kinetic energy very 
rapidly.

In region 2, the trajectory is very close to the $x$--axis 
where the scalar field energy density is kinetic energy dominated
(hence $y \approx 0$), and
its behaviour in the $x$ direction is described by
\begin{equation}
\label{KEdom} x'=-{3\over2}(2-\gamma)x(1-x^2) \ .
\end{equation}  
For $x<0$, $x'$ is always positive and $x$ increases; for $x>0$, 
$x'$ is always negative and $x$ decreases bringing the trajectory 
towards the $\lambda$ -- axis (in the $3$--dimensional phase space 
($x,y,\lambda$)).

Close to the $\lambda$--axis, the system is described by
\begin{eqnarray}
x' = -{3\over2}(2-\gamma)x \,, \hspace{1cm} 
y' = \frac{3}{2}\gamma y \,.
\end{eqnarray}
Therefore, the trajectory in region 3 evolves along the $x$ 
direction at a rate of $x \propto a^{-3(2-\gamma)/2}$. Since
the background fluid is dominant at this stage, 
$H^2 \propto a^{- 3 \gamma}$, we have the kinetic energy falling off
rapidly as
$\dot{\phi}^2 \sim x^2 H^2 \propto a^{-6}$. The trajectory moves in
the $y$ direction as $y \propto a^{3\gamma/2}$, that is 
$V \sim y^2 H^2 \approx \mbox{constant} $ and the scalar field 
is essentially ``frozen''.

%%%%%%%%%%%%%%%%%%%%%%%%%%%%%%%%%%%%%%%%%%%%%%%%%%%%%%
\section{ The tracker solution}
%%%%%%%%%%%%%%%%%%%%%%%%%%%%%%%%%%%%%%%%%%%%%%%%%%%%%%
\label{tracker}
In this section we study the case in which the scalar field
evolution
is well approximated by a linear relation between $x$ and $y$, as it
happens
along region 4 of the example illustrated in Fig.~\ref{fig2}.
In order to
study the evolution of the scalar field when the slope of the potential
is varying, {\it i.e.} $\lambda$ is changing with time, we rewrite
Eqs. (\ref{dynamics}) by defining        
$\epsilon=1/\lambda$, $x = \epsilon X$, and $y = \epsilon Y$.

We then have,
\begin{eqnarray}
\label{backsystem}
X' &=& \sqrt{6}(\Gamma-1)X^2 -3X + \sqrt{\frac{3}{2}}Y^2 + \, \nonumber \\
   &~& \frac{3}{2}X[2\epsilon^2X^2 + 
       \gamma(1-\epsilon^2X^2-\epsilon^2Y^2)] \nonumber \,, \\
Y' &=& \sqrt{6}(\Gamma-1)X Y -\sqrt{\frac{3}{2}}X Y + \, \\ 
   &~& \frac{3}{2}Y[2\epsilon^2X^2 +
       \gamma(1-\epsilon^2X^2-\epsilon^2Y^2)] \nonumber \,, \\ 
\epsilon' &=& \sqrt{6} \epsilon (\Gamma-1) X \,. \nonumber
\end{eqnarray}
For $\epsilon$ small ($\lambda$ large) or $\Gamma\approx1$, 
$\epsilon$ becomes nearly constant. Moreover, if $\Gamma$ 
is also nearly constant  we can solve $X'=Y'=0$ and
find the ``instantaneous critical points'',
\begin{eqnarray}
\label{critical1}
x_c(\lambda) = \sqrt{\frac{3}{2}} \frac{\gamma_{\phi}}{\lambda} \,, \hspace{1cm}
y_c^2(\lambda) = \frac{3}{2} \frac{\gamma_{\phi}}{\lambda^2}(2-\gamma_{\phi}) \,, 
\end{eqnarray}
where the equation of state of the scalar field is
\begin{eqnarray}
\label{eqstate}
\gamma_{\phi} &=&  \frac{1}{2} [\gamma + (2\Gamma -1)\lambda^2/3] \pm
\nonumber \, \\
              &~& \frac{1}{2} \sqrt{ \left[ - \gamma +
                      (2\Gamma-1)\lambda^2/3 \right]^2 
                      + 8\gamma(\Gamma-1)\lambda^2/3 } \,.
\end{eqnarray}
%
%\begin{eqnarray}
%\label{eqstate}
%\gamma_{\phi} &=&  \frac{\gamma}{2} + (2\Gamma-1)\frac{\lambda^2}{6} 
%                   \pm \nonumber \, \\
%              &~&  \sqrt{ \left[ \frac{\gamma}{2} + 
%                    (2\Gamma-1)\frac{\lambda^2}{6} \right]^2 - 
%		   \gamma \frac{\lambda^2}{3}} \,.
%\end{eqnarray}
%
%\begin{eqnarray}
%\label{eqstate}
%\gamma_{\phi} &=&  \frac{1}{2} \left(\gamma + (2\Gamma-1)\lambda^2/3 \right) 
%                    \times \nonumber \, \\
%              &~&     \left[ 1 \pm \sqrt{ 1 - \frac{4}{3} 
%                   \frac{\gamma \lambda^2}{(\gamma + (2\Gamma
%                    -1)\lambda^2/3 )^2 }} ~~\right] \,.
%\end{eqnarray}
%

The plus root leads to unphysical results, so only the negative one will
be used in what follows. The contribution of the field to the total
energy density is $\Omega_{\phi} = 3 \gamma_{\phi}/\lambda^2$.

Note that in the limit when $\Gamma-1 \approx 0$ one has
%
%\begin{equation}
%\gamma_{\phi} =  \gamma \left[1+ \frac{2(\Gamma-1)}{2\Gamma-1 
%                                   -3\gamma/\lambda^2}   \right]^{-1}   \,.
%\end{equation}
%
\begin{equation}
\label{eqstate2}
\gamma_{\phi} = \gamma \left[1- \frac{2(\Gamma-1)}{1 
                                   -3\gamma/\lambda^2}   \right]   \,.
\end{equation}
In other words, for potentials with small curvature, the equation of
state of the scalar field is very close to the equation of state of
the background fluid, and it is said that the field ``tracks'' the
background fluid.
This expression can account for values of $\Omega_{\phi} > 1/2$.

From Eq. (\ref{eqstate}), it can also be shown that in the limit of
large $\lambda$,
when the background fluid is completely dominating,
we recover the expression in \cite{stein},
\begin{equation}
\label{steinhardt}
\gamma_{\phi} = \frac{\gamma}{2\Gamma-1} .
\end{equation}
In the limit when $\Gamma-1 \approx 0$, the above expression is equivalent
to Eq. (\ref{eqstate2}) in the limit of large $\lambda$.

One can now study the stability of the ``instantaneous critical points''
linearising Eqs. (\ref{backsystem}) in $u$ and $v$ with 
$X = X_c + u$ and $Y= Y_c + v$ resulting in the first-order equations
of motion,
\begin{eqnarray}
\left( \begin{array}{c} u' \\ v'\end{array} \right) = \mathcal{M}
\left( \begin{array}{c}  u \\ v \end{array} \right) .
\end{eqnarray}
Using Eq. (\ref{eqstate}) to write $\Gamma-1$ in terms of
$\gamma_{\phi}$, we find the eigenvalues,
\begin{eqnarray}
\label{eigenv1}
m_{\pm} = -\frac{3}{4\lambda^2}\left[ (\gamma-\gamma_{\phi})
            (3\gamma_{\phi}+\lambda^2) + (2-\gamma_{\phi})\lambda^2
            \right] \times   \nonumber \, \\
          \left[ 1 \pm \sqrt{ 1- \frac{ 8~\lambda^2(2-\gamma_{\phi})
              (\gamma \lambda^2-3\gamma_{\phi}^2)}{
              [ (\gamma - \gamma_{\phi})(3\gamma_{\phi} + \lambda^2) + 
              (2- \gamma_{\phi})\lambda^2 ]^2}} ~\right] \,.
\end{eqnarray}
It is straightforward to check that Eq. (\ref{eigenv1}) reduces to
Eq. (A8) of \cite{copeland} when $\gamma = \gamma_{\phi}$ ({\it i.e.} 
the pure exponential case) and to Eq. (15) of \cite{stein}
in the limit of large $\lambda$.

\begin{figure}[ht!]
\includegraphics[height=6cm,width=8cm]{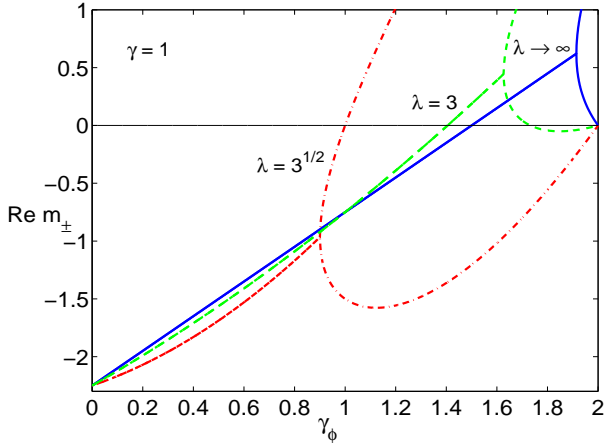}
\caption{ \label{fig3} Dependence of the real part of the eigenvalues on 
$\gamma_{\phi}$ and $\lambda$ for a background fluid of matter. Upper
and lower lines represent respectively solutions $-$ and $+$ in
Eq. (\ref{eigenv1}). }
\end{figure}

\begin{figure}[ht!]
\includegraphics[height=6cm,width=8cm]{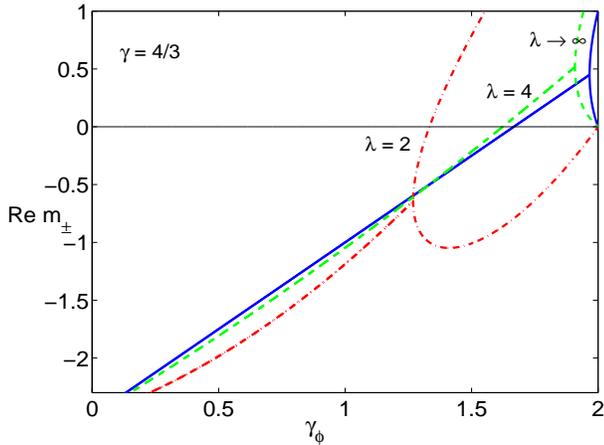}
\caption{ \label{fig4} Dependence of the real part of the eigenvalues on 
$\gamma_{\phi}$ and $\lambda$ for a background fluid of radiation. Upper
and lower lines represent respectively solutions $-$ and $+$ in
Eq. (\ref{eigenv1}). } 
\end{figure}

The system is stable if the real part of both eigenvalues is negative.
From Eq. (\ref{eigenv1}), this is completely assured if,
\begin{eqnarray}
\label{gfalso}
\gamma_{\phi} &<& \sqrt{\frac{\gamma \lambda^2}{3}} \nonumber
                  \,, {~~~ \rm and}   \\    
\gamma_{\phi} &<& \frac{3\gamma-2\lambda^2}{6} \left(
                  1- \sqrt{ 1 + \frac{12 \lambda^2 (2+\gamma)}
                  {(3\gamma-2\lambda^2)^2}} ~\right ) \,.
\end{eqnarray}
Moreover, if the quantity under the square root of  Eq. (\ref{eigenv1})
is negative, the critical points are a stable spiral; and a stable
node, otherwise.
In Fig. \ref{fig3} and Fig. \ref{fig4} one can see the dependence of
the eigenvalues on $\gamma_{\phi}$ and $\lambda$ for a background
fluid of matter 
($\gamma = 1$) and radiation ($\gamma = 4/3$), respectively. The lines
split when the quantity under the square root of  Eq. (\ref{eigenv1})
becomes positive. The first of the conditions in Eqs. (\ref{gfalso})
marks the point for which only one of the real parts of the 
eigenvalues becomes
positive, and the second condition the point where both real parts of
the eigenvalues
become positive. In the limit of large $\lambda$ the latter is 
$\gamma_{\phi} = 1+ \gamma/2$.


For a large class of scalar potentials $V(\phi)$,
stability is possible for a  
large number of $e$-folds allowing a scalar 
field sub-dominance during a long period of time as we will see in 
the applications below. We should point out that so far these results 
are general since we have not assumed, yet, any type of potential 
$V(\phi)$.
               
%%%%%%%%%%%%%%%%%%%%%%%%%%%%%%%%%%%%%%%%%%%%%%%%%%%%%%%%%
\section{The scalar field dominated solution}
%%%%%%%%%%%%%%%%%%%%%%%%%%%%%%%%%%%%%%%%%%%%%%%%%%%%%%%%%
\label{field}
We now consider the case in which the evolution of the scalar field 
approaches the unit circle $x^2+y^2 = 1$, as it is the case along region 5 of
Fig. 2.
Introducing the new variables
$x = \lambda X$ and $y^2 = 1- \lambda^2 Y^2$, Eqs. (\ref{dynamics}) give,
\begin{eqnarray}
\label{fieldsystem}
X' &=& \sqrt{6}\lambda^2 (\Gamma-1)X^2 -3X +
              \sqrt{\frac{3}{2}}(1-\lambda^2 Y^2)^2 + \nonumber \, \\
          &~& \frac{3}{2}\lambda^2X[2X^2+\gamma(-X^2 + Y^2)] \nonumber
              \,, \\ 
YY' &=& \sqrt{6}\lambda^2(\Gamma-1)X Y^2 + \sqrt{\frac{3}{2}}X
              (1-\lambda^2 Y^2) - \nonumber \, \\
    &~& \frac{3}{2}(1-\lambda^2 Y^2)[2 X^2+\gamma(- X^2 + Y^2)] \,.
\end{eqnarray}

However, the same kind of analysis as for the tracker solution becomes
here extremely complicated.
We take then a set of simple reasonable assumptions in order to obtain simple
and useful results.
                
One can see from Fig. \ref{fig2} that at late times, when the scalar 
field is dominant, $y$ is approximately $1$ and $x$ approaches
zero. This means the scalar potential is overtaking the energy density and 
the potential is very flat. 
In other words, $\lambda$ is getting closer to zero. Let us take then
$ \lambda \approx 0$ and $\lambda^2(\Gamma-1)$ nearly constant.
Eq. (\ref{fieldsystem}) then reads,
\begin{eqnarray}
X' &=& \sqrt{6}\lambda^2(\Gamma-1)X^2 -3X + \sqrt{\frac{3}{2}}
\nonumber \,, \\
YY' &=&  \sqrt{6}\lambda^2(\Gamma-1) XY^2 + \sqrt{\frac{3}{2}}X -
\nonumber \, \\
    &~& \frac{3}{2} [2X^2 + \gamma(-X^2-Y^2)] \,.
\end{eqnarray}
The system has critical points in,
\begin{eqnarray}
\label{critical2}
x_c(\lambda) = \frac{\lambda_{\phi}}{\sqrt{6}} \,, \hspace{1cm}
y_c^2(\lambda) = 1 - \frac{\lambda_{\phi}^2}{6}\,.
\end{eqnarray}
Hence, the scalar field is dominant, $\Omega_{\phi} = 1$ and 
$\gamma_{\phi} = \lambda_{\phi}^2/3$, where
we have defined
\begin{equation}
\label{lambdaphi}
\lambda_{\phi} = \frac{3}{2} \left[\frac{1 
                 \pm \sqrt{1-4(\Gamma-1)\lambda^2/3}}
                                     {(\Gamma-1)\lambda}\right] , 
\end{equation}
for $\Gamma \neq 1$, and $\lambda_{\phi} = \lambda$ otherwise.
As before, only the minus solution has physical meaning.
Expanding $\lambda_{\phi}$ we can approximate the solution by
\begin{equation}
\label{lfi}                                      
\lambda_{\phi} = \lambda \left[1+\frac{1}{3}(\Gamma-1)\lambda^2 \right] \,. 
\end{equation}

As we did for the tracker solution, we perturb the solutions 
around the critical points to study their stability. Expanding
Eq. (\ref{fieldsystem}) and using Eq. (\ref{lambdaphi}) to write 
$\Gamma -1 $ in terms of $\lambda$ and $\lambda_{\phi}$, we find 
the following eigenvalues:
\begin{eqnarray}
\label{eigenv2}
m_{+} &=& 6~(\lambda_{\phi}-\lambda)\frac{1}{\lambda_{\phi}}
          + \frac{1}{2}(\lambda_{\phi}^2 + \lambda \lambda_{\phi}
          - 6\gamma)  \nonumber \,, \\
m_{-} &=& 3 - 6~\frac{\lambda}{\lambda_{\phi}} + \frac{1}{2} 
              \lambda_{\phi}(3 \lambda_{\phi} -2 \lambda) \,.
\end{eqnarray}
 
\begin{figure}[ht!]
\includegraphics[height=6cm,width=8cm]{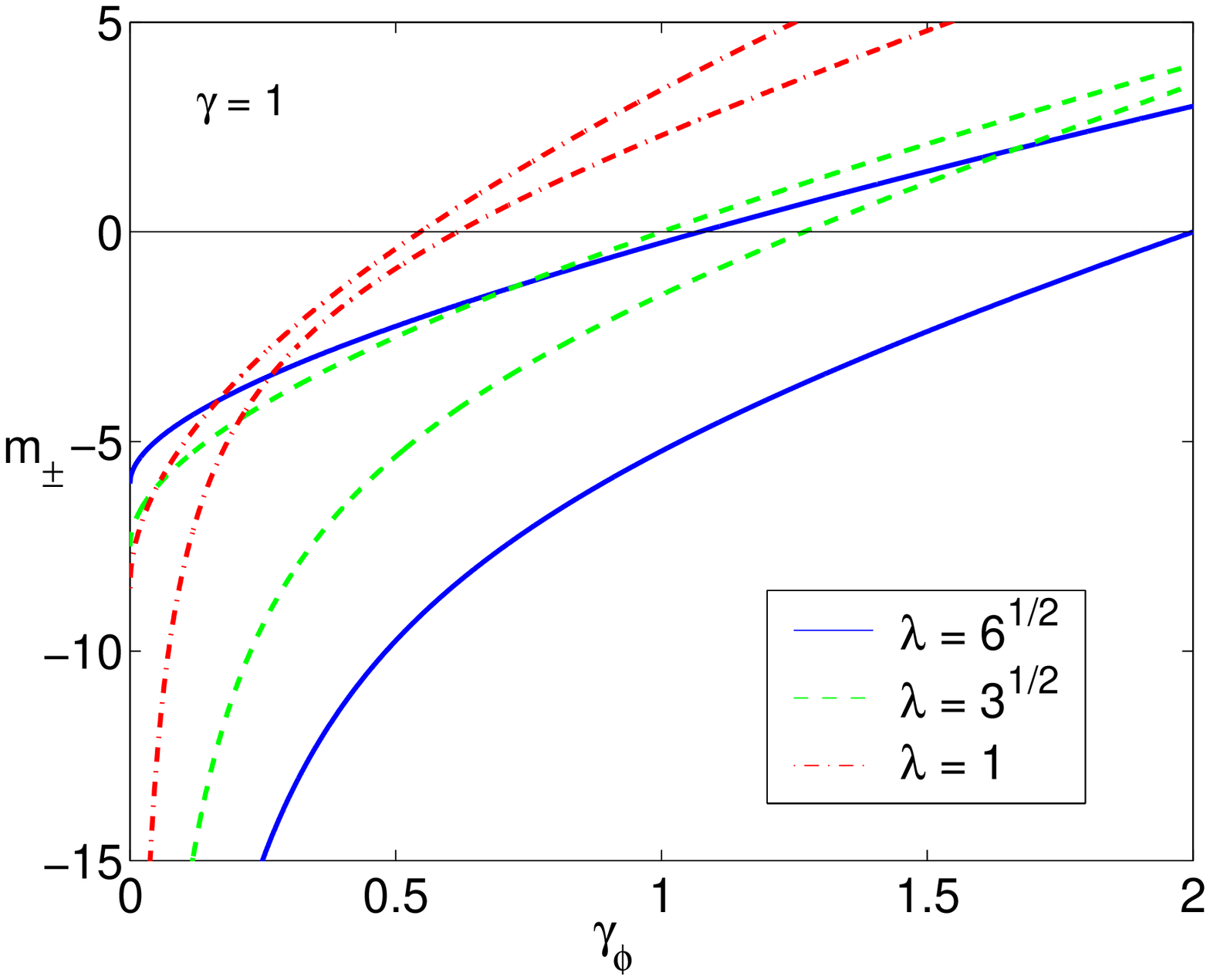}
\caption{ \label{fig5} Dependence of the eigenvalues on 
$\gamma_{\phi}$ and $\lambda$ for a background fluid of matter. 
Upper and lower lines represent respectively 
solutions $m_{+}$ and $m_{-}$ in Eq. (\ref{eigenv2}). }
\end{figure}

\begin{figure}[ht!]
\includegraphics[height=6cm,width=8cm]{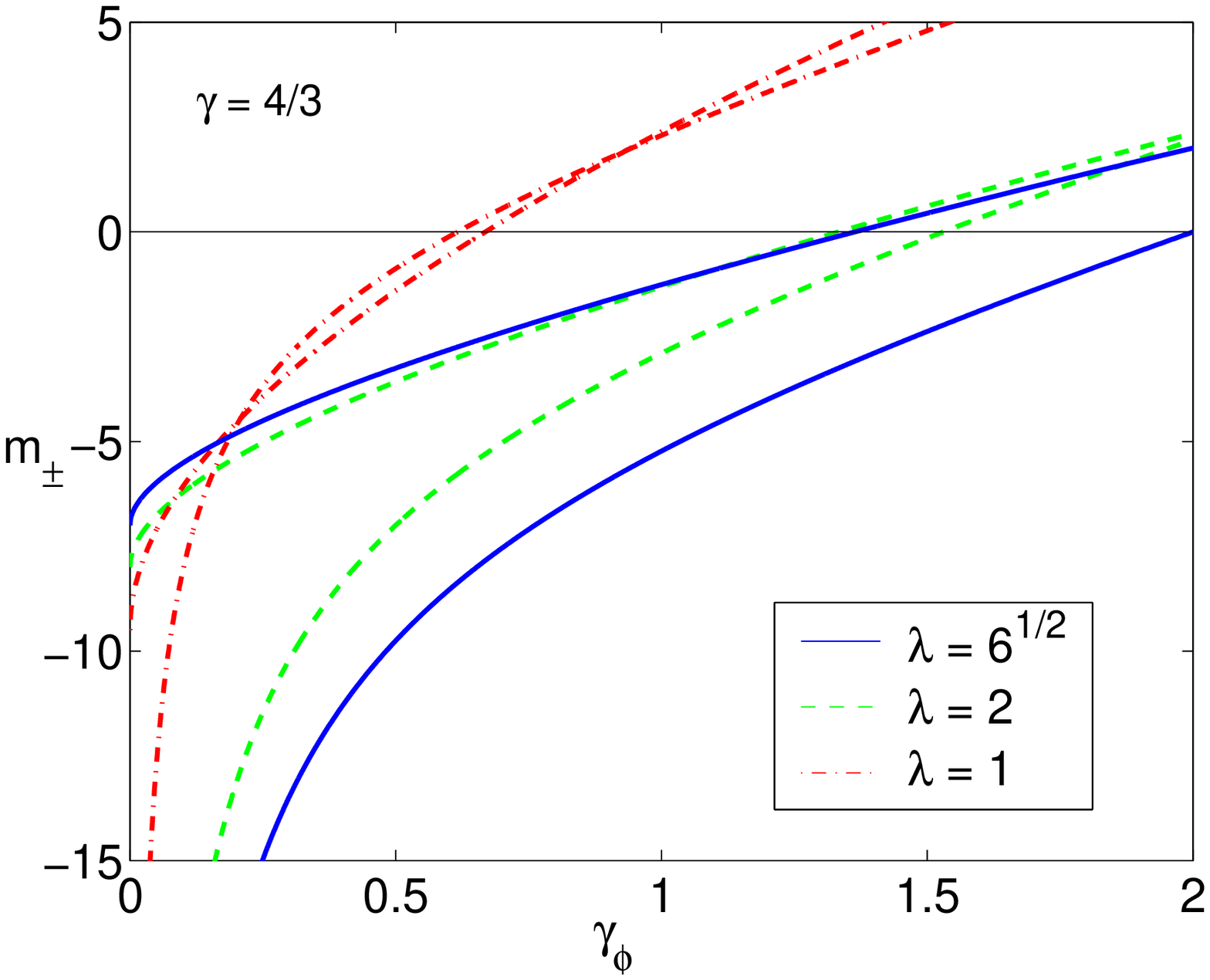}
\caption{ \label{fig6} Dependence of the eigenvalues on 
$\gamma_{\phi}$ and $\lambda$ for a background fluid of
radiation. Upper and lower lines represent respectively 
solutions $m_{+}$ and $m_{-}$ in Eq. (\ref{eigenv2}). }
\end{figure}

For $\Gamma =1$ this expression reduces to 
the eigenvalues found in \cite{copeland}, as we would expect. 
In Fig. \ref{fig5} and Fig. \ref{fig6} we show the dependence of 
the eigenvalues on $\gamma_{\phi}$ and $\lambda$ for a background
fluid of matter and radiation, respectively.

For completeness, we indicate that the background fluid dominated solution with 
$(x,y,\lambda) = (0,0,\lambda)$ has eigenvalues,
\begin{eqnarray}
m_{+} = \frac{3\gamma}{2}\,, \hspace{1cm} 
m_{-} = - \frac{3(2-\gamma)}{2}\,,
\end{eqnarray}
and when $\Gamma = 1$, there exist the kinetic dominated solutions 
$(x,y,\lambda) = (\pm 1,0,\lambda)$ with eigenvalues,
\begin{eqnarray}
m_{+} =  3~(2-\gamma) \,, \hspace{0.5cm} 
m_{-} =  \sqrt{\frac{3}{2}}~(\sqrt{6} \mp \lambda)   \,.
\end{eqnarray}

In Table \ref{tabela} we give a summary of the properties of
these critical points.

%%%%%%%%%%%%%%%%%%%%%%%%%%%%%%%%%%%%%%%%%%%%%%
%               TABELA              
%%%%%%%%%%%%%%%%%%%%%%%%%%%%%%%%%%%%%%%%%%%%%%
\begin{table*}[t]
\begin{center}
\begin{tabular}{|c|c|c|c|c|c|}
{$x_c$} & {$y_c^2$} & {Existence} & {Stability} & {$\Omega_{\phi}$} 
& {$\gamma_{\phi}$}\\ \hline
$0$ & $0$ & all $\gamma$ and $\lambda$ & saddle point for $0<\gamma<2$ & $0$ 
     & undefined \\
$1$ & $0$ & all $\gamma$ and $\lambda$ 
     & unstable node for $\lambda < \sqrt{6} $,~ 
       saddle point for  $\lambda > \sqrt{6} $ &  $1$ & $2$ \\
$-1$ & $0$ & all $\gamma$ and $\lambda$ 
     & unstable node for $\lambda > -\sqrt{6} $,~ 
       saddle point for  $\lambda < -\sqrt{6} $ & $1$ & $2$ \\       
$\sqrt{\frac{3}{2}} \frac{\gamma_{\phi}}{\lambda}$ 
     & $\frac{3}{2} \frac{\gamma_{\phi}}{\lambda^2}(2-\gamma_{\phi})$
     & $\lambda^2 > 3\gamma_{\phi}$ & stable for $Re (m_{\pm}) < 0$ 
       (see Eq. (\ref{eigenv1})) 
     & $\frac{3\gamma_{\phi}}{\lambda^2}$ & Eq. (\ref{eqstate}) \\
$\frac{\lambda_{\phi}}{\sqrt{6}}$ & $1 - \frac{\lambda_{\phi}^2}{6}$  
     & $\lambda_{\phi}^2 < 6$ & stable for $m_{\pm} < 0$ 
       (see Eq. (\ref{eigenv2})) 
     & $1$ & $\frac{\lambda_{\phi}^2}{3}$ \\
\end{tabular}
\end{center}
\caption{\label{tabela} The properties of the critical points. 
$\lambda_{\phi}$ is defined in Eq. (\ref{lambdaphi}). }
\end{table*}

%%%%%%%%%%%%%%%%%%%%%%%%%%%%%%%%%%%%%%%%%%%%%%%%%
\section{Applications}
%%%%%%%%%%%%%%%%%%%%%%%%%%%%%%%%%%%%%%%%%%%%%%%%%
\subsection{Quintessence}
The first estimate for the value of the cosmological constant from
current particle physics is the Planck scale, while the lowest is
set by the electroweak scale, namely, of order $10^8 ~{\rm GeV}^4$.
These are tremendously high values on cosmological energy scales.
So, the simplest solution so far has been to assume that by some mechanism, 
the cosmological constant would vanish altogether. 

In the last decade, measurements of the apparent magnitude--redshift 
relation using SnIa combined with CMB and galaxy clusters and light 
elements abundances measurements, give indications that we
are living in an accelerating Universe with 
$\Omega_{\rm matter} \sim 0.3$ and $\Omega_{\Lambda} \sim
2/3$ \cite{data}. 
Then, the discovery that the cosmological constant like term
is small but non zero, is disturbing. Having today a cosmological constant
energy density contribution, $\rho_{\Lambda} = \Lambda/8 \pi G$, of
the same order of magnitude as the critical energy density 
$\rho_c \sim 10^{-47} {\rm GeV}^4$, requires to fine tune its
initial value to 120 orders of magnitude below the Planck scale. 
In order to alleviate this puzzle, the idea of ``quintessence'' was
introduced \cite{zlatev}. 
One of its versions consistes of an inhomogeneous
scalar field $Q$
rolling down a potential with attractor solutions. The argument is
that if the scalar field joins the attractor solution before the
present epoch, information about the initial conditions will get lost,
which allows a freedom in choosing those conditions within 100 orders of
magnitude, thus relieving the fine tuning issue.
The initial suggestion for such potentials was to use an inverse power law form 
$ V(Q) \propto Q^{-\alpha}$ which can be found in models of supersymmetric QCD
\cite{binetruy}.
Pure exponential potentials $V \propto \exp(\lambda Q)$ also have attractor 
solutions; however, they cannot be used on their own to model quintessence
\cite{scale}
but interesting modifications have been proposed such as 
$V \propto \exp(-\lambda Q)Q^{-\alpha}$ \cite{skordis},
sum of pure exponentials $V \propto \exp(\alpha Q) + \exp(\beta Q)$
\cite{nelson}, Supergravity
inspired models with $V \propto \exp(Q^2)Q^{-\alpha}$ \cite{brax} or
$V \propto \exp(Q^{4/3})Q^{2/3}$  \cite{rosati}.

In this section we will give the scalar field and its equation of state 
evolution in a background fluid dominated setting, for the inverse power 
law potential. We will then turn to the general potential 
$V \propto \exp(\alpha Q^{\beta}) Q^{\mu}$.

Consider the inverse power law potential,
\begin{equation}
V(Q) = \frac{M^{4+\alpha}}{Q^{\alpha}} \,.
\end{equation}
for which the relevant quantities are
\begin{eqnarray}
\lambda = -\frac{\alpha}{\kappa Q}, ~~~~~ \Gamma-1 = \frac{1}{\alpha} \,.
\end{eqnarray}
We have seen in section 
\ref{tracker} that when the background fluid is dominant the equation
of state of the scalar field can be well approximated by 
Eq. (\ref{steinhardt}). Therefore, in this case the equation of state is a 
constant and is given by 
\begin{equation}
\label{eqspower}
\gamma_{Q} = \frac{\gamma \alpha}{2+\alpha} .
\end{equation}
Integrating the $x$ equation in Eq. (\ref{critical1})
the solution for the scalar field reads,
\begin{equation}
\label{qpower}
Q = Q_i \left(\frac{a}{a_i}\right)^{3\gamma/(2+\alpha)},
\end{equation}
where $Q_i$ can be derived from the $y$ equation assuming 
$H^2 \approx \kappa^2/3~ \rho_B^i \exp(-3 \gamma N)$, 
with $N = \ln(a/a_i)$, to give
\begin{equation}
Q_i = \left[\frac{3}{2} \frac{\kappa^2 \rho_B^i}{M^{4+\alpha}} 
       \frac{\gamma_{Q}(2-\gamma_{Q})}{\alpha^2}\right]^{-1/(2+\alpha)} .
\end{equation}

We review here some of the properties of power law type of
potentials \cite{scherrer}. For example, Eq.(\ref{eqspower}) still
holds for negative $\alpha$ provided $\alpha < -2$. Moreover, we have
seen that for large $\lambda$, the solutions are stable provided 
$\gamma_{Q} < 1+ \gamma/2$, which yields,
\begin{eqnarray}
\alpha &>& 2 \frac{\gamma+2}{\gamma-2}, \hspace{1cm} \alpha > 0
\nonumber \,, \\
\alpha &<& 2 \frac{\gamma+2}{\gamma-2}, \hspace{1cm} \alpha < 0 \,. 
\end{eqnarray}
For $\alpha > 0$ the condition is always true, however, for $\alpha <
0$ it imposes the bounds $\alpha < -6$ and $\alpha < -10$ for matter and
radiation dominated fluids, respectively.
For even, negative $\alpha$ the field can have early oscillatory
behaviour with average equation of state given by the virial theorem
\begin{equation}
<\gamma_{Q}> = 1+ \frac{\alpha + 2}{\alpha -2 },
\end{equation} 
while energy is continually being converted between kinetic and potential.

\begin{figure}[ht!]
\includegraphics[height=6cm,width=8cm]{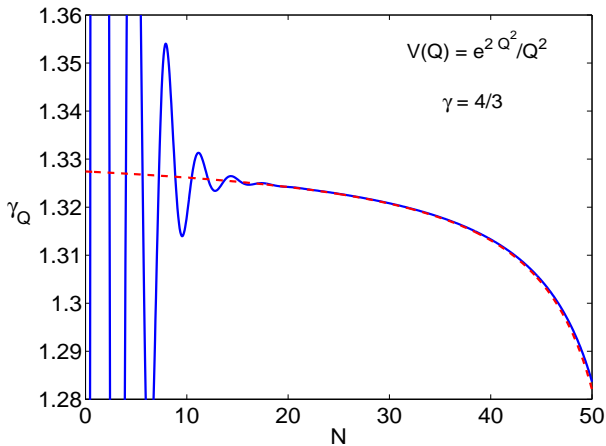}
\caption{ \label{fig7} Evolution of the equation of state of the
scalar field in a radiation background. The solid line is the
numerical evolution and the dashed line our approximate
analytical solution using Eqs. 
(\ref{sol1}), (\ref{lagama}) and (\ref{steinhardt}).}
\end{figure}

Consider now the more general potential,
\begin{equation}
V(Q) = M^4 e^{\alpha (\kappa Q)^{\beta}} (\kappa Q)^{\mu} \, ;
\label{genpot}
\end{equation}
we have 
\begin{eqnarray}
\label{lagama}
\lambda &=& -( \alpha \beta (\kappa Q)^{\beta} + \mu ) (\kappa Q)^{-1} \nonumber \, ,\\
\Gamma -1 &=& \frac{\alpha \beta (\beta-1) (\kappa Q)^{\beta} -\mu }
                 {(\alpha \beta (\kappa Q)^{\beta} + \mu)^2}
\end{eqnarray}
and integrating $x$ in Eq. (\ref{critical1}) with $\gamma_{\phi}$
given by Eq. (\ref{steinhardt}), we find the solution
\begin{eqnarray} 
\label{solucao1}                
\alpha (\kappa Q)^{\beta} &=& \Phi_i+ 
          \frac{2-\mu}{\beta} \ln(\kappa Q)^{\beta} \nonumber \, \\ 
                         &-& 2 \ln [ \alpha \beta(\kappa Q)^{\beta} + \mu ] -
                             3 \gamma N \, ,
\end{eqnarray}
where the last logarithm term only exists if $\alpha \neq 0$.
$\Phi_i$ is an integration constant and can be derived 
from the $y$ equation. If we assume $\kappa Q \gg \mu/\alpha \beta$ then
we have a nearly constant $\Gamma-1 \approx 0$, and we can neglect this 
contribution from the equation of state, to yield the simple result,

\begin{equation}
\label{bigfi}
\Phi_i =  \ln \left[ \frac{3}{2} 
           \frac{\rho_B^i}{M^4} \gamma (2-\gamma) \right] .
\end{equation}         
Since we do not have an explicit expression for $(\kappa Q)^{\beta}$
we take the ansatz used in \cite{barreiro98a}, which is a perturbative
solution,
\begin{eqnarray}
\label{sol1}
\alpha (\kappa Q)^{\beta}_1 &=& \Phi_i  - 3 \gamma N \nonumber \,, \\ 
\alpha (\kappa Q)^{\beta}_n &=& \Phi_i +
                     \frac{2-\mu}{\beta} \ln(\kappa Q)^{\beta}_{n-1} 
                           \nonumber\, \\
                     &-&
                     2 \ln [\alpha \beta (\kappa Q)^{\beta}_{n-1}+ \mu] -
                     3 \gamma N  ,
\end{eqnarray}                                                   
Substituting this solution back in  Eqs. (\ref{lagama}) and (\ref{steinhardt}) we have now
the complete evolution of the scalar field equation of state in terms
of the background fluid energy density and model parameters, only.
We can see from Fig. \ref{fig7} that the second order solution
is, in general, already a good approximation.                      

Note that for a pure exponential ($\beta =1$, $\mu = 0$), we recover
the exact solution
\begin{equation}
Q = Q_i - \frac{3 \gamma}{\kappa \alpha} N,
\end{equation}
where $Q_i = \Phi_i/\kappa \alpha$.

When $\kappa Q \ll \mu/\alpha \beta$, a positive definite
potential,  Eq. (\ref{genpot}) becomes well approximated by a pure power 
law potential, 
{\it i.e.}
$\Gamma-1 \approx -1/\mu$, $\alpha \approx 0$ and, neglecting the
second logarithm contribution,
we rewrite Eq. (\ref{solucao1}) as,
\begin{eqnarray}
\kappa Q =\Psi_i ~\left( \frac{a}{a_i} \right)^{3 \gamma/(2-\mu)} \,,
\end{eqnarray}           
where $\Psi_i = \exp(\Phi_i/(\mu-2))$, is estimated, as before, 
using the $y$ equation to yield,
\begin{equation}
\Psi_i = \left[\frac{3}{2} \frac{\rho_B^i}{M^4}\frac{1}{\mu^2}
               \frac{\gamma \mu}{\mu-2} \left(2-\frac{\gamma \mu}{\mu-2} \right) \right]^
               {-1/(2-\mu)} \,.
\end{equation}
Despite the above solution being merely an approximation, substituting
back in Eq. (\ref{lagama}), the variation of the equation of state with
time can be well accounted for.
%
%%%
%
%
%
\subsection{Moduli stabilisation}
In string theory, the modulus  and
the dilaton (moduli) play an important
role. They parametrise the structure of the compactified manifold and
their vacuum expectation value (vev) determine the $d=4$ value of the
gauge and gravitational couplings and fix the unification scale
$M_{\rm GUT}$. 

Stabilisation of the moduli fields in the $d=4$
effective theory, has been studied by including nonperturbative effects
such as multiple gaugino condensates, which develop a minimum, and 
non-perturbative corrections to the K\"{a}hler
potential \cite{choi}. Unfortunately, the scalar potentials in these
theories are exponentially
steep, therefore, it is expected that these fields roll past the
minimum rather than acquiring a vev \cite{brustein}. Hence,
stabilisation of the moduli is one of the most serious questions in
string theory. Solutions to this problem have been suggested in the literature.
One possibility is that matter fields other than the moduli drive the 
evolution of the Universe \cite{barreiro98a,huey}. If this is the
case, it has been shown that this setting opens up a wider region of the
parameter space for which dynamical stabilisation of the moduli is
successful. The reason behind this feature is, once again, the
existence of attractor solutions that slow down the fields by the
amount needed to trap them when they reach the minimum. 

Away from the minimum, the evolution of the moduli can be written in
terms of a canonically normalised field $\sigma$, in a double
exponential potential.

We take the general potential
\begin{equation}
V(\sigma) = M^4 \exp(\alpha e^{\beta \kappa \sigma}),
\end{equation}
where $\alpha$ is an arbitrary real number.
Following the same line of argument as before, we find the solution
for $S \equiv \alpha \exp(\beta \kappa \sigma)$ to be (in accordance
with \cite{barreiro98a}), 
\begin{eqnarray}
\label{sol2}
S_1 &=& \Phi_i - 3 \gamma N \nonumber \,, \\
S_n &=& \Phi_i - 2 \ln (S_{n-1}/\alpha) - 3 \gamma N ,
\end{eqnarray}
and $\Phi_i$ is  given by 
\begin{equation}
\label{bigfi2}
\Phi_i =  \ln \left[ \frac{3}{2} 
           \frac{\rho_B^i}{M^4} \frac{\gamma (2-\gamma)}
           {\alpha^2 \beta^2} \right] .
\end{equation}    

This attractor solution enables us to stabilise the dilaton at the minimum.
One sees that the equation of state of the scalar field is the same
as that of the background fluid, apart from a logarithmic deviation and an upper
or lower shift given by $\Phi_i$. In other 
words, the ratio between kinetic and potential energy of the scalar field is
roughly a constant and hence prevents the field from rolling past the 
minimum for certain values of the background fluid equation of state.

A last remark is that if $\alpha$ is positive, then $\lambda$ is decreasing
and this becomes a realistic quintessence potential, as well.

\subsection{Quintessential inflation}
We have seen in section \ref{field} that for small enough values of
$\lambda_{\phi}$ the scalar field drives the dynamics of the
Universe. In this section we present scalar field dominated solutions 
for the same potentials used above. The particular case of
$\lambda_{\phi} < 2$ is of extreme importance, since it corresponds to
an inflationary scenario ($\ddot{a} \propto -\rho_{\phi}-3p_{\phi} >
0$). In what follows we consider a good approximation, taking
$\lambda_{\phi} \approx \lambda$.

Considering the limit $(\kappa Q)^{\beta} \gg \mu/\alpha \beta$, we
integrate $x$ in Eq.(\ref{critical2}) to obtain for the
potential $V(Q)= M^4 \exp(\alpha (\kappa Q)^{\beta})(\kappa Q)^{\mu}$ the
solution, 
\begin{equation}
(\kappa Q)^{\beta}  = \left[ (\kappa Q_i)^{2-\beta} + \alpha \beta
(2-\beta) N \right]^{\beta/(2-\beta)} \,, 
\end{equation}
when $\beta \neq 2$, and
\begin{equation}
(\kappa Q)^2  = -\frac{\mu}{\alpha \beta} + \left[ (\kappa Q_i)^2 +
\frac{\mu}{\alpha \beta} \right] \left(\frac{a}{a_i}\right)^{-2\alpha \beta} \,, 
\end{equation}
otherwise (see also \cite{barrow}). The case $(\kappa Q)^{\beta} \ll
\mu/\alpha \beta$ 
can be well approximated by the power law solution,
\begin{equation}
(\kappa Q)^{\beta}  = \left[ (\kappa Q_i)^2 - 2\mu N \right]^{\beta/2} \,. 
\end{equation}

From Eq. (\ref{lagama}), it should be clear that for 
$(\kappa Q)^{\beta} \gg \mu/\alpha \beta$, if $\beta < 1$, 
$\lambda$ is increasing and we can imagine a scenario in which the
Universe is first inflating and then, when $\lambda$ is such that $\lambda^2
> 2$ the universe starts to decelerate. Reheating can happen by gravitational 
particle production by which conventional particles are created
quantum mechanically from the time varying gravitational field 
\cite{gravpp}.
Moreover, if we choose negative $\mu$, the potential has a minimum
with non vanishing vacuum energy. Assuming this minimum is the
responsible for the currently accelerated expansion of the universe,
we can envisage a scenario of ``quintessential inflation''
\cite{quinfl}.

Taking the number of {\it e}-folds from the end of inflation to be
$N = 50$ and imposing the spectral index to be $ n_{\rm S} > 0.95$, a
closer look at this potential reveals that 
\begin{equation}
\beta < 0.01,
\end{equation}
and for the approximation $(\kappa Q)^{\beta} \gg -\mu/\alpha\beta$ to
be valid,
\begin{equation}
-0.1 \lesssim \mu < 0.
\end{equation}
The amplitude of density perturbations $A_{\rm S} = 2 \times 10^{-5}$
measured by COBE, then imposes
\begin{eqnarray}
\alpha >  293.8 ~, \hspace{1cm}
M < 10^{-17} {\rm GeV}.
\end{eqnarray}

The ratio of tensorial to scalar perturbations is then, 
$A_{\rm T}^2/A_{\rm S}^2 < 0.17$, and the tensorial spectral index 
$n_{\rm T} > -0.028$.

An important condition in order not to spoil nucleosynthesis is that
the value of the Hubble constant when inflation ends must be 
$H_{\rm end} > 10^{8} {\rm GeV}$ which is amply satisfied in this model as 
$H_{\rm end} > 10^{12} {\rm GeV}$.

For the double exponential potential, $V(\phi) = M^4 \exp(\alpha e^{\beta
\kappa \phi})$ the field rolls according to
\begin{equation}
\phi = -\frac{1}{\kappa \beta} \ln \left( e^{-\beta \kappa \phi_i} -
\alpha \beta^2 N \right) \,.
\end{equation}
When $\alpha$ is negative the potential offers a plateau for 
$\beta \phi < 0$, where inflation can happen. Assuming that particles
are produced gravitationally as above, to satisfy the value of
density perturbations we find, 
$M \approx 10^{15} {\rm GeV}$ and $n_{\rm S} \approx
0.96$. Furthermore, for $|\beta| > 1$, these numbers are nearly independent 
of the precise values of $\alpha$ and $\beta$.
We must also require, $\beta > 0.1$ in order not to spoil 
nucleosynthesis predictions.
As noted in \cite{macorra}, even if the field would reach the tracker
solution before today, its contribution to the total energy density 
would be ever decreasing. So the scalar field can never explain, in
this case, the present acceleration, unless some other contribution is
added to the potential  to
make the scalar field become dominant today.

Although interesting, most of  these models of quintessential inflation
lead to a regime where the kinetic energy is the dominant contribution
to the
scalar field energy density until its value is of order the critical
energy density today. Therefore, the contribution of the scalar field
at the time of recombination is negligible and the value of the field
equation of state today is very close to the one of the cosmological
constant, $\gamma_{\Lambda} = 0$. 
Hence, it is not expected that CMB and SnIa observations will be
able to provide direct evidence for these scenarios. A second problem 
associated with these models is the overproduction of gravitinos, as
the reheating temperature $T_{\rm RH}$ is often 
$T_{\rm RH} \sim H_{\rm end} \gtrsim 10^9 ~{\rm GeV}$, leading to dangerous cosmological 
consequences \cite{gravitino}. And of course, a solid theoretical
motivation for the suggested phenomenological potentials is still 
to be found.

\section{Summary}
We have studied the dynamics of a scalar field in a flat FRW Universe
with a barotropic fluid. We have considered the general case of field 
dependent slope $\lambda$ of the scalar potential and analysed the
differences, especially in the stability conditions,  with
respect to the constant slope case. These results are
summarised in table \ref{tabela}. 
We exemplified the tracker and scalar field dominated solutions by
giving explicit calculations of the field evolution for different
potentials. We have discussed applications of these solutions to the
issues of quintessence, moduli stabilisation and quintessential
inflation. Finally, we would like to emphasise that we have
demonstrated here the existence and stability of a number of scalar
attractor solutions which have been previously only assumed in the literature.

\acknowledgements
We acknowledge conversations with Martin Eriksson, David Mota, 
Luis Ure\~{n}a-L\'{o}pez and David Wiltshire. We are grateful to Tiago
Barreiro, Dominic Clancy, Ed Copeland and Andrew Liddle for comments 
on the manuscript. S.C.C.N. and F.R. would like to thank for the 
kind hospitality the staff of the Centre
for Theoretical Physics at University of Sussex, where this work was
begun.
N.J.N. is supported by
Funda\c{c}\~{a}o para a Ci\^{e}ncia e a Tecnologia (Portugal).

\end{document}